\documentclass[prd,singlecolumn,amsmath,nobibnotes,nofootinbib,superscriptaddress,preprintnumbers]{revtex4}

\usepackage{bm} \usepackage{graphicx}
\usepackage{epstopdf}
\def\ba{\begin{eqnarray}}
\def\ea{\end{eqnarray}}
\def\be{\begin{equation}}
\def\ee{\end{equation}}
\def\({\left(}
\def\){\right)}
\def\[{\left[}
\def\]{\right]}
\def\<{\left<}
\def\>{\right>}
\def\Q{Q}

\begin{document}

\title{Extremal limits and black hole entropy}
\date{\today}
\author{Sean M. Carroll}
\affiliation{California Institute of Technology, Pasadena, CA 91125, USA}
\author{Matthew C. Johnson}
\affiliation{California Institute of Technology, Pasadena, CA 91125, USA}
\author{Lisa Randall}
\affiliation{Harvard University, Cambridge, MA 02138, USA}

\begin{abstract}
Taking the extremal limit of a non-extremal Reissner-Nordstr\"om black hole (by externally varying the mass or charge), the region between the inner and outer event horizons experiences an interesting fate -- while this region is absent in the extremal case, it does not disappear in the extremal limit but rather approaches a patch of $AdS_2\times S^2$. In other words, the approach to extremality is not continuous, as the non-extremal Reissner-Nordstr\"om solution splits into two spacetimes at extremality: an extremal black hole and a disconnected $AdS$ space. We suggest that the unusual nature of this limit may help in understanding the entropy of extremal black holes.
\end{abstract}

\preprint{CALT-68.2717}

\maketitle

\section{Introduction}
\label{sec:introduction}

In this note, we investigate a little-appreciated feature of the classical black hole geometry that distinguishes extremal black holes from their near-extremal cousins. In addition to the black hole solutions, it has long been known that in Einstein-Maxwell theory with or without a cosmological constant, it is possible to find static solutions that are the product of maximally symmetric spaces~\cite{Nariai:1950uq,Bertotti:1959pf,Robinson:1959fk}. If the background cosmological constant is zero, the relevant solution takes the form of $AdS_2\times S^2$, two-dimensional anti-de~Sitter times a two-sphere of constant radius. These ``compactification solutions'' differ from the black hole solutions by boundary conditions. 

Beginning with a non-extremal black hole and considering the limit of extremality, it was noted by~\cite{Ginsparg:1982rs} that the standard static coordinate system becomes pathological. Surprisingly, the region between degenerating horizons remains of constant four-volume as the limit is taken. In this note we will review how this region, together with a region just outside the horizon, forms the compactification solution. Considering regions a finite proper distance away from the horizon, and then taking the limit, one obtains an extremal black hole. Thus, even at the level of classical geometry, the extremal limit is discontinuous. Subtleties in the limit can be important in general. In this note we focus on the implications for the entropy of extremal Reissner-Nordstr\"om black holes.

Black holes have long been an important theoretical laboratory for exploring the nature of quantum gravity. The fact that black holes radiate~\cite{Hawking:1974sw} and exhibit a formal similarity to thermodynamical systems~\cite{Christodoulou:1970wf,Hawking:1971tu,Bardeen:1973gs} leads to the association of a Bekenstein-Hawking entropy to the black hole, proportional to the area of its event horizon~\cite{Bekenstein:1973mi,Hawking:1974sw}. While there has been great success in reproducing the Bekenstein-Hawking entropy formula using a variety of  methods (see e.g.~\cite{wald-2001-4}), the theory of black hole entropy is still incomplete. 

One apparent inconsistency seems to arise when considering extremal black holes using semiclassical methods~\cite{Gibbons:1976ue}, which seem to indicate the extremal black holes have vanishing entropy even when the area of the event horizon is non-zero~\cite{Hawking:1994ii,Teitelboim:1994az}\footnote{Extremal black holes are distinguished from non-extremal black holes at the classical level as well -- they cannot be produced  by a process involving any finite number of steps without violating the weak energy condition~\cite{Bardeen:1973gs}. (Modes of formation might also account for the entropy~\cite{Zurek:1985gd}, see also~\cite{Pretorius:1997wr}.) This is in accord with the view that extremal black holes are to be thought of as solitons, which are typically expected to be formed only quantum mechanically by pair production (and whose entropy one expects to vanish)~\cite{Gibbons:1994ff}.}. However, it is exactly in the extremal cases that string theory microstate counting was first used to calculate a non-zero entropy~\cite{Strominger:1996sh}, matching what is expected from the Bekenstein-Hawking formula (see e.g.~\cite{Horowitz:1996qd,Peet:2000hn} for reviews). 

A third way to determine the black-hole entropy, complementary to semiclassical techniques and string-theory microstate counting, may be referred to as ``dual microstate counting.''  Central to this approach is the observation that the near-horizon geometry of an extremal black hole is locally that of two-dimensional anti-de~Sitter space cross a two sphere.  The symmetries of this geometry define a conformal field theory (CFT), the entropy of which can be determined and associated with the black hole~\cite{Strominger:1997eq}. This method does not rely on supersymmetry, makes no reference to string theory, and has recently been applied to charged, spinning black holes in four and higher dimensions~\cite{Guica:2008mu,Hartman:2008pb,Lu:2008jk}.

The fundamental reason for the discrepancy between the string theory and dual microstate counting results and the semiclassical calculation remains elusive, largely because of a lack of precise overlap between the semiclassical and string theory methods; see~\cite{Horowitz:1996qd,Das:1996rn,Ghosh:1996gp,Horowitz:2005vp,Zaslavsky:1996zz,Zaslavsky:1997ha,Silva:2006xv,Dias:2007dj}  for possible  resolutions. The semiclassical calculations attempt to infer information about the microphysics from the classical geometry, while the string theory methods attempt to infer something about the classical geometry from the microphysics.

We suggest the above observation about the discontinuous nature of the extremal limit might  allow us to shed some light on the issue of black hole entropy discussed above. Since the limit includes the $AdS$ solution as well as the extremal black hole, the additional space might account for the net entropy. The dual microstate counting calculation relies on a holographic dual picture on the horizon and does not make explicit reference to the full black hole geometry. Taking into account both of the classical solutions in the limit also yields the correct entropy, but the the origin of the $AdS$ space in this picture is quite different, comprising a portion of the non-extremal spacetime in the extremal limit as opposed to the dual near-horizon region of an extremal black hole extended to a full $AdS$ space.

In Sec.~\ref{sec:geometry}, we describe the black hole solutions, and in Sec.~\ref{sec:extremal limit} we carefully examine the limiting procedure.  In Sec.~\ref{sec:uniqueness} we verify that the spacetimes we are considering, Reissner-Nordstr\"om and $AdS_2 \times S^2$, are the only static, spherically symmetric solutions of the Einstein-Maxwell system.  We conclude by discussing possible implications of the discontinuous extremal limit for questions of black hole entropy in Sec.~\ref{sec:discussion}. 

\section{Black hole and compactification solutions}\label{sec:geometry}

In this section we consider the properties of four-dimensional, static, spherically symmetric solutions to the Einstein-Maxwell system with zero cosmological constant (in units where $G=1$),
\begin{equation}\label{eq:action}
S = \frac{1}{16 \pi} \int d^4 x \sqrt{-g} \left( R - \frac{F^2}{4} \right).
\end{equation}
One set of solutions is the Reissner-Nordstr\"om black hole, with metric
\begin{equation}\label{eq:rnmetric}
ds^2 = - \frac{(r - r_{+}) (r - r_{-})}{r^2} dt^2 + \frac{r^2}{(r - r_{+}) (r - r_{-})} dr^2 + r^2 d\Omega_2^2
\end{equation}
and field strength
\begin{equation}\label{eq:fieldstrength}
F = \frac{Q_e}{r^2} dt \wedge dr + Q_m \sin \theta d\theta \wedge d\phi.
\end{equation}
This set of coordinates does not cover the entire manifold, and there are event horizons located at the coordinate singularities 
\begin{equation}
r_{\pm} =  M \pm \sqrt{M^2 - \Q^2},
\end{equation}
where we have defined
\be
  \Q \equiv \sqrt{Q_e^2 + Q_m^2}
\ee
(which we take to always be positive).
Exactly at extremality, where $M = \Q$, the event horizons coincide at the extremal radius:
\be
\rho \equiv r_+ = r_- = M = \Q.
\ee
For masses less than this, the spacetime possesses a naked timelike singularity; we will not consider such solutions. The extremal black hole metric is given by
\begin{equation}\label{eq:extremalbh}
ds^2 = - \frac{(r - \rho)^2}{r^2} dt^2 + \frac{r^2}{(r - \rho)^2} dr^2 + r^2 d\Omega_2^2.
\end{equation}

\begin{figure*}
\begin{center}
\includegraphics[height=8cm]{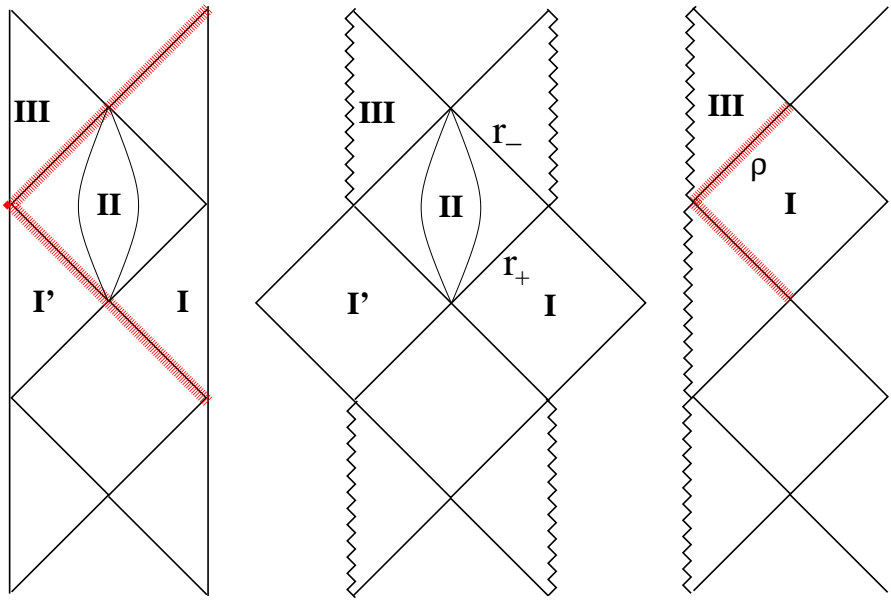}
\end{center}
\caption{The causal structure of the $AdS_2$ space (left), non-extremal Reissner Nordstr\"om black hole (center), and the extremal Reissner Nordstr\"om black hole (right). The non-extremal black hole possesses two event horizons at $r = r_+$ and $r = r_-$, while the extremal black hole possesses only one at $r=\rho$. The solid lines in Region~II of the non-extremal black hole are timelike trajectories of constant $\psi$ (see Eq.~\ref{eq:regionIIcoords}) extending from $r_+$ to $r_-$. The horizon of the extremal black hole solution indicated by the hatched red line is locally equivalent to the hatched red line of the $AdS_2$ diagram. }
\label{fig:rnads}
\end{figure*}

The causal structure of the (extremal and non-extremal) Reissner-Nordstr\"om black hole is shown in Fig.~\ref{fig:rnads} (along with that of anti-de~Sitter space). There are three different types of patches, labeled as follows:
\be
\begin{array}{lc}
{\rm Region ~I}: & r_+ < r < \infty, \ \ \ \ -\infty < t < \infty  \cr 
{\rm Region ~II}: & r_- < r < r_+, \ \ \ \ -\infty < t < \infty \cr 
{\rm Region ~III}: & 0 < r < r_-, \ \ \ \ -\infty < t < \infty \, .
\end{array}
\ee
The metric Eq.~\ref{eq:rnmetric} will cover each region separately.  Note that in Region~II between the horizons, it is $r$ rather than $t$ that plays the role of a timelike coordinate, and the geometry is that of a homogeneous space with geometry ${\bf R}\times S^2$.  The spherical part of the geometry contracts monotonically in ``time'' (from $r_+$ to $r_-$) while the ${\bf R}$ part expands from zero ``scale factor'' and then re-contracts to zero.

Another set of solutions to the same Einstein-Maxwell system is the product of a maximally extended $AdS_2$ and a stabilized sphere $S^2$ of the same radius $\rho$. The metric
\begin{equation}\label{eq:adsmetric}
ds^2 = \frac{\rho^2}{\cos^2 \theta} \left( -d \tau^2 + d\theta^2 \right) + \rho^2 d \Omega_2^2,
\end{equation}
with
\begin{equation}
\rho = \Q,
\end{equation}
covers the entire manifold, with the coordinate ranges $-\frac{\pi}{2} \leq \theta \leq \frac{\pi}{2}$ and $-\infty \leq \tau \leq \infty$. These are the compactification solutions. Note that $AdS_2$ has two causally separate timelike boundaries at $\theta = \pm \pi / 2$.  The full causal structure of the $AdS_2$ quotient is shown in Fig.~\ref{fig:rnads}.

Both the extremal black holes and compactification solutions can be specified by the set of charges $\{ Q_e, Q_m \}$ since, from the extremality condition,  the mass $M$ of the extremal black hole is no longer an independent parameter. The two types of solutions are distinguished by the imposed boundary conditions on the two-sphere (as we discuss in more detail in Sec.~\ref{sec:uniqueness}); the charges alone (together with the assumption that the solution is static and possesses spherical symmetry) are not sufficient to specify the global properties of the solution.

However, there is a sense in which the extremal and compactification solutions are {\em locally} equivalent in the near-horizon limit. Changing the spacelike radial coordinate in Eq.~\ref{eq:extremalbh} to
\begin{equation}
\lambda = \frac{r - \rho}{\rho} ,
\end{equation}
the metric becomes
\begin{equation}
ds^2 = - \frac{\lambda^2}{\left( 1+\lambda \right)^2} dt^2 + \frac{ \left( 1+\lambda \right)^2}{\lambda^2} d\lambda^2 +\rho^2 \left(  1+\lambda \right)^2 d\Omega_2^2 .
\end{equation}
A peculiar property of the extremal metric is that the proper distance along a $t = \rm{const.}$ slice from any point $\lambda$ to the horizon at $\lambda_0 \rightarrow 0$ is logarithmically divergent
\begin{equation}
\int_{\lambda_0}^{\lambda} d\lambda \frac{1+\lambda}{\lambda} = \lambda - \lambda_0 + \log \left( \frac{\lambda}{\lambda_0} \right).
\end{equation}

Taking the near-horizon limit of the extremal black hole, $\lambda \rightarrow 0$, the metric becomes
\begin{equation}
ds^2 = - \lambda^2 dt^2 + \frac{\rho^2}{\lambda^2} d \lambda^2 + \rho^2 d\Omega_2^2, 
\end{equation}
which can be recognized as $AdS_2 \times S^2$ after transforming to the coordinates in Eq.~\ref{eq:adsmetric}
\begin{equation}
t = \frac{\rho \sin \tau}{\cos \tau - \sin \theta}, \ \ \ \ \lambda = \frac{\cos \tau - \sin \theta}{\cos \theta}.
\end{equation}
From this relation, it can be seen that the horizon at $\lambda = 0$ is identified with $\theta= \pm \tau + \pi/2$. This is highlighted in Fig.~\ref{fig:rnads} by the red hatched line. The extremal solution has regions near the horizon that locally approximate $AdS_2 \times S^2$, but the approximation becomes exact only at the location of the horizon, which is an infinite proper distance from any point in the exterior of the black hole. 

\section{The extremal limit}\label{sec:extremal limit}

It is possible to obtain both the extremal black hole and compactification solutions by taking various limits of a non-extremal black hole. We will work at the level of the fully extended classical geometry, and  will consider Regions~I--III in turn. 

We begin with Region~II of the non-extremal solution where $r_- < r < r_+$ ($r$ is a timelike coordinate in this region). In the limit of extremality $r_{\pm} \rightarrow \rho$, it would appear that this region is continuously diminished to zero size. However, because the metric coefficients diverge on either side of the range in $r$, this need not be true. Following Refs.~\cite{Ginsparg:1982rs,Bousso:1996au,Mann:1995vb,Dias:2003up}, we can see this more clearly by defining 
\be
r_- = \rho - \epsilon, \ \ \ r_+ = \rho + \epsilon, 
\ee
so that $\rho = \Q$ is the value of $r$ to which the two horizons evolve, while $\epsilon = \sqrt{M^2 - \Q^2}$ parameterizes the deviation from extremality. In the following, we will hold $\rho$ fixed, while varying $\epsilon$. Then we can define a new timelike coordinate $\chi$ and spacelike coordinate $\psi$ via
\begin{equation}\label{eq:regionIIcoords}
r = \rho - \epsilon \cos \chi, \ \ \ \psi = \frac{\epsilon}{\rho^2} t .
\end{equation}
These coordinates allow us to zoom in on the near-horizon region. They range over $0 < \chi < \pi$ and $-\infty < \psi < \infty$, and the horizons become degenerate in the limit where $\epsilon \rightarrow 0$. The metric is given by
\begin{equation}\label{eq:fullinsideepsmetric}
ds^2 = \rho^2 \left[ - \left( 1 - \frac{\epsilon}{\rho} \cos \chi \right)^2 d\chi^2 + \frac{\sin^2 \chi}{\left( 1 - \frac{\epsilon}{\rho} \cos \chi \right)^2} d\psi^2 + \left( 1 - \frac{\epsilon}{\rho} \cos \chi \right)^2 d\Omega_2^2 \right]  .
\end{equation}

We first consider  what happens to the spacetime volume of Region~II as we approach extremality. To investigate this, we consider the spacetime distance between the inner and outer event horizons.  This is equivalent to the proper time elapsed on a trajectory of constant finite $\psi$ (such trajectories are shown in Fig.~\ref{fig:rnads}) between $0 < \chi < \pi$.  (If we fix one point on a spacelike hypersurface, the distance from that point to some other hypersurface is determined by a curve of maximum proper time; in the present context it is straightforward to verify that a curve of constant finite $\psi$ satisfies this criterion, and that the result is independent of the initial point chosen.) This proper time is given by
\begin{equation}
\Delta \tau = \rho \int_{0}^{\pi} d\chi  \left( 1 - \frac{\epsilon}{\rho} \cos \chi \right) = \pi \rho .
\end{equation}
Remarkably, this is independent of $\epsilon$, the deviation from extremality. So, even in the limit where the radii of the horizons become coincident, Region~II between them does not vanish; the inner and outer event horizons remain a constant physical distance apart.

Taking the limit $\epsilon \rightarrow 0$, the metric of Region~II becomes
\begin{equation}\label{eq:insideepsmetric}
ds^2 = \rho^2 \left[ - d\chi^2 + \sin^2 \chi d\psi^2 +  d\Omega_2^2 \right]  ,
\end{equation}
which is a portion of $AdS_2 \times S^2$ with a sphere of constant radius $\rho$. The portion of the $AdS_2$ that is covered can be determined by going to the global coordinates of Eq.~(\ref{eq:adsmetric}),
\begin{equation}
\cos \chi  = \frac{\cos \tau}{\cos \theta} , \ \ \ \ \ \tanh \psi = \frac{\sin \theta}{\sin \tau}.
\end{equation}
Over the full range in $\{ \chi, \psi \}$, Region~II of the $AdS_2$ in Fig.~\ref{fig:rnads} is filled out. 

Note that we obtained this portion of $AdS_2\times S^2$ as the extremal limit was approached by starting from Region~II, in between the inner and outer horizons. This region is separated from the asymptotic boundary conditions at spatial infinity of the black hole solution, and the ``black-hole-ness'' disappears entirely from this solution at the extremal point. 

We now turn to Region~I, where the choice of different spacetime regions and asymptotic boundary conditions will play a role. A new set of coordinates can be introduced that cover all of Region~I, 
\begin{equation}\label{eq:outsidecoords}
r_- = \rho - \epsilon, \ \ \ r_+ = \rho + \epsilon, \ \ \ r = \rho + \epsilon \cosh \chi, \ \ \ \psi = \frac{\epsilon}{\rho^2} t.
\end{equation}
Here $\chi = 0$ at $r = r_+$ and $\chi = \infty$ at future/past null infinity, and $-\infty < \psi < \infty$. The metric in these coordinates is  
\begin{equation}
ds^2 = \rho^2 \left[ - \frac{\sinh^2 \chi }{\left( 1+ \frac{\epsilon}{\rho} \cosh \chi  \right)^2} d\psi^2 + \left( 1+ \frac{\epsilon}{\rho} \cosh \chi  \right)^2 d \chi^2 + \left( 1+ \frac{\epsilon}{\rho} \cosh \chi  \right)^2 d \Omega_2^2 \right] .
\end{equation}

Because $\cosh \chi$ can grow large enough at $\chi \rightarrow \infty$ to compensate for $\epsilon \rightarrow 0$ in Eq.~\ref{eq:outsidecoords}, it is important to establish the spacetime location of interest before taking the extremal limit. If $\chi$ is set to a fixed finite value and then $\epsilon$ is sent to zero, we are effectively taking the near-horizon and extremal limit simultaneously since from Eq.~\ref{eq:outsidecoords}, $r \rightarrow \rho$. However, if we fix $r \neq \rho$, and then send $\epsilon \rightarrow 0$ (with $\chi$ becoming commensurately large), we are taking only the extremal limit. The near horizon limit discussed in Sec.~\ref{sec:geometry} is taken after $\epsilon = 0$, in the exactly extremal geometry.

For fixed finite $\chi$ and $\epsilon \rightarrow 0$ the metric approaches
\begin{equation}\label{eq:outsideadsmetric}
ds^2 = \rho^2 \left[ - \sinh^2 \chi d\psi^2 + d\chi^2 + d\Omega_2^2 \right].
\end{equation}
This describes a  piece of $AdS_2 \times S^2$. The coordinate transformation to the global $AdS_2$ coordinates is in this case
\begin{equation}
\cosh \chi = \frac{\cos \tau}{\cos \theta}, \ \ \ \ \tanh \psi = \frac{\sin \tau}{\sin \theta},
\end{equation}
where Region~I of the $AdS_2$ space in Fig.~\ref{fig:rnads} is filled out over the full range in $\{\chi, \psi\}$. Performing a similar limiting procedure in Region~III of the non-extremal black hole solution, Region~III of the $AdS$ solution (together with the timelike boundary) is produced. Combining these patches with the patch covered by Eq.~\ref{eq:insideepsmetric}, the fully extended $AdS$ solution is generated in the limit, with the timelike boundary arising from the portions of Regions~I and III just outside/inside of the horizon. 

If we look at fixed finite $r \neq \rho$, then clearly the coordinates Eq.~\ref{eq:outsidecoords} become inappropriate at $\epsilon = 0$. In this case, it is more appropriate to use the original $\{ t, r \}$ coordinates of Eq.~\ref{eq:rnmetric}. Taking the limit where $r_+ = r_-$, we obtain the extremal black hole metric Eq.~\ref{eq:extremalbh}, which covers Region~I of the extremal black hole solution in Fig.~\ref{fig:rnads}. A similar procedure can be applied to Region~III, which yields the interior of the extremal black hole. These two patches, at extremality, provide a global cover of the extremal black hole solution. 

In the extremal black hole geometry, it can be shown that the past and future event horizons never intersect~\cite{Hawking:1973uf}. In the non-extremal black hole geometry, there is an intersection occurring at  $\chi = 0$. Taking $\epsilon \rightarrow 0$ at finite $\chi$, this intersection is preserved, and by the limiting procedure described above, becomes part of the $AdS_2 \times S^2$ space. Therefore, it is clear that the non-extremal black hole exterior does in fact have two regions that become distinctly different spacetimes in the limit of extremality. 

\begin{figure*}
\begin{center}
\includegraphics[height=8cm]{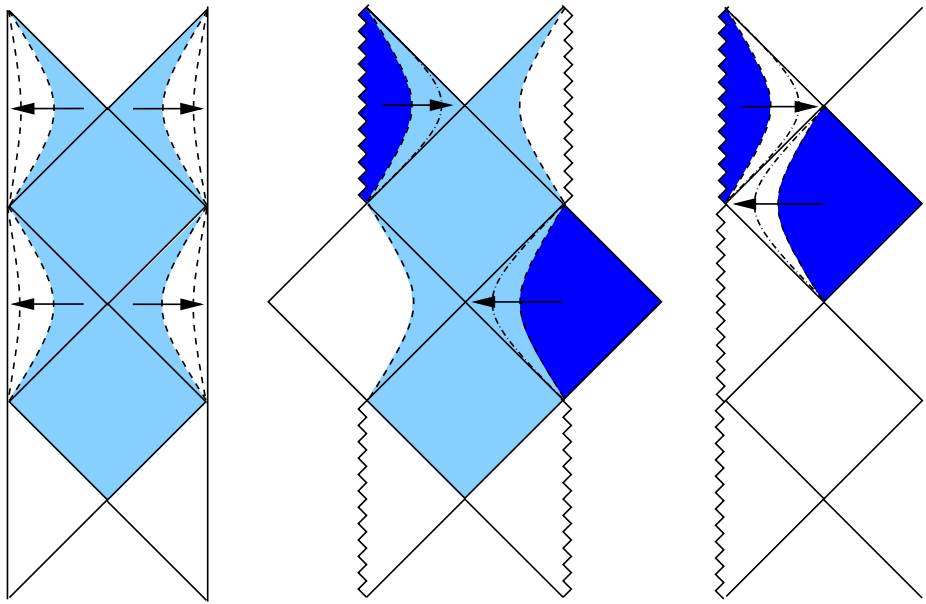}
\end{center}
\caption{A pictorial representation of the limiting procedure. The $AdS_2 \times S^2$ space (left) and extremal black hole (right) can be obtained from different regions of the non-extremal black hole (center). For fixed constant $r^* \sim \rho$ (dashed lines in the non-extremal black hole diagram), regions with smaller $r$ (the light shaded portions of the diagram comprising Region~II and portions of Regions~I and III) are approximated close to extremality by the corresponding light shaded regions of the $AdS_2 \times S^2$ diagram. These interfaces approach the timelike boundaries of the $AdS_2 \times S^2$ space when extremality is approached ($\epsilon \rightarrow 0$), as indicated by the arrows. The dark shaded regions on the non-extremal black hole diagram are approximated close to extremality by the corresponding dark shaded regions on the extremal black hole diagram. As extremality is approached, the extremal black hole   approximation applies closer and closer to the horizon (as indicated by the arrows).}
\label{fig:rnadsshaded}
\end{figure*}

In Fig.~\ref{fig:rnadsshaded} we depict the limiting process. In the center panel is the non-extremal black hole. For a black hole near extremality ($r_+ \sim \rho$), if we choose a fixed radius $r = r^*$ near the outer horizon in Region~I (indicated by the dashed lines), this will correspond to a fixed value of $\chi^* \sim \log \left[ \left( r^* - \rho \right) / \epsilon \right]$ from the relation Eq.~\ref{eq:outsidecoords}. We can also define an equivalent fixed radius inside of Region~III. In the light shaded portions of the non-extremal black hole solution bounded by these radii, the radius of the $S^2$ will be approximately constant, and the metric will locally approximate the light shaded portion of the $AdS_2 \times S^2$ space shown in the left panel. Fixing $r^*$ and taking $\epsilon$ smaller, the value of $\chi^*$ will increase, as indicated by the arrows in the left panel. If we take $r^* \rightarrow \rho$ as $\epsilon \rightarrow 0$, we recover a timelike boundary of the $AdS$ solution (rendering the $\{ \chi, \psi \}$ coordinates ill-defined outside of the black hole, as described above). Including Regions~II and III, the global $AdS_2 \times S^2$ compactification solution is obtained.

Shown in the right panel of Fig.~\ref{fig:rnadsshaded} is the extremal solution. Close to extremality, the portions of the non-extremal black hole outside of $r^*$ will approximate the extremal solution more closely than the compactification solution, since the size of the $S^2$ will nowhere be approximately constant. The dark shaded region of the non-extremal black hole will map onto the dark shaded regions of the extremal solution in Fig.~\ref{fig:rnadsshaded}. As extremality is approached, this region will grow, as indicated by the arrows, until at $\epsilon = 0$, the full extremal black hole is obtained.

So, we have seen that in the extremal limit, Region II and the near--vicinity of the horizon in Regions I and III of the non--extremal black hole become the compactification solution, while the portions of Region I and III any finite distance away from the horizon form the extremal black hole. In Regions I and III, it is important to distinguish the order of near--horizon and extremal limits to determine if a portion of the extremal black hole or a portion of the compactification solution is reached at extremality ($\epsilon = 0$ exactly). 

The discussion above generalizes to arbitrary dimension for the Einstein-Maxwell system with a $(D-2)$-form field strength, and to the case where there is a non-zero cosmological constant. Adding a non-zero positive cosmological constant changes the horizon structure of the solutions, since for a range of charges, there will be a cosmological horizon in addition to the inner and outer black hole event horizons. With non zero cosmological constant, two compactification solutions can be generated by the same limiting procedure of degenerating horizons described above, where a portion of $dS_2 \times S^{D-2}$ space is formed between degenerating outer black hole and cosmological horizons, and a portion of $AdS_2 \times S^{D-2}$ space is formed between degenerating inner and outer black hole horizons.

\section{Uniqueness}\label{sec:uniqueness}

According to the black hole uniqueness theorems \cite{Israel:1967za,Heusler:1998ua}, 
the Reissner-Nordstr\"om solution is the unique spherically symmetric, asymptotically
flat, static (where static is defined with respect to the asymptotically flat region) solution to the Einstein-Maxwell equations. It is clear that if we drop the assumption of asymptotic flatness, another solution is allowed:  the compactification solution $AdS_2\times S^2$, which has different boundary conditions. In this section we verify that this is the \emph{only} new solution that arises upon dropping the assumption of asymptotic flatness.

Assuming only that the spacetime is spherically symmetric and possesses a static region, the metric can be written as 
\begin{equation}\label{eq:staticansatz}
ds^2 = - A(z)^2 dt^2 + dz^2 + r(z)^2 d\Omega_2^2.
\end{equation}
This is fully general, although these coordinates will typically only cover some portion of the full solution. The components of the Einstein tensor are
\begin{eqnarray}
G_{tt} &=& -\frac{A^2}{r^2} \left( {r'}^2 + 2 r r''  -1 \right) \\
G_{zz} &=& \frac{1}{A r^2} \left( 2 r {A'} {r'} + A {r'}^2 - A \right) \\
G_{\theta \theta} &=& \frac{r}{A} \left( A' r' +r A'' + A r''\right) \\
G_{\phi \phi} &=& \sin^2 \theta G_{\theta \theta},
\end{eqnarray}
while the components of the energy-momentum tensor for the field strength Eq.~\ref{eq:fieldstrength} are given by
\begin{eqnarray}
T_{tt} &=& A^2  \frac{Q^2}{8 \pi r^4} \\
T_{zz} &=& - \frac{1}{A^2} T_{tt} \\ 
T_{\theta \theta} &=& \frac{r^2}{A^2} T_{tt} \\
T_{\phi \phi} & = & \frac{r^2 \sin^2 \theta}{A^2} T_{tt}.
\end{eqnarray}

We will first look for static solutions where $r'=r''=0$ for all $z$. From the $tt$ and $zz$ Einstein equation, we obtain the constant value of $r$,
\begin{equation}
r_0^2 = \rho^2.
\end{equation}
From the $\theta \theta$ and $\phi \phi$ equations, we obtain an equation for $A(z)$,
\begin{equation}\label{eq:Aeom}
A'' = \frac{1}{\rho^2} A.
\end{equation}
Choosing $A(0)=0$, $A'(0) = \rho^{-1}$ yields
\begin{equation}
A(z) = \sinh (z / \rho),
\end{equation}
and the metric Eq.~\ref{eq:staticansatz} can be recognized as a rescaled version of the portion of $AdS_2 \times S^2$ covered by Eq.~\ref{eq:outsideadsmetric}. It is possible to analytically continue $z \rightarrow i z$, and extend the coordinates across $z=0$ where $A=0$ to the portion of $AdS_2 \times S^2$ covered by Eq.~\ref{eq:insideepsmetric}. Continuing to extend the coordinates across points where $A = 0$ in each region, yields global $AdS_2 \times S^2$, and thus the entire compactification solution.

\begin{figure*}
\begin{center}
\includegraphics[height=5cm]{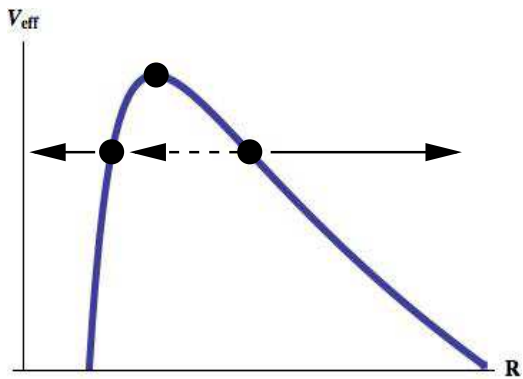}
\end{center}
\caption{The effective potential Eq.~\ref{eq:veffec}.  There is a solution that sits precisely at
the maximum, corresponding to $AdS_2 \times S^2$.  All other solutions correspond to 
some part of the RN spacetime; energies higher than the maximum of $V_{eff}$ are
super-extremal, equal to the maximum are extremal, and below the maximum are
sub-extremal.  The sub-extremal region between the horizons at $r_+$ and $r_-$ 
corresponds to motion in an inverted potential between two turning points.}
\label{fig:rnbhtrajectories}
\end{figure*}

We now look for solutions where $r$ is a function of $z$. Combining the $tt$ and $zz$ Einstein equations, we ascertain that
\begin{equation}\label{eq:Aadnrp}
A = r'.
\end{equation}
From the $tt$ Einstein equation, we obtain an equation of motion for $R$
\begin{equation}\label{eq:Reom}
r'' + \frac{{r'}^2}{2 r} = \frac{1}{2r} \left( 1- \frac{Q^2 }{r^2} \right) = - \frac{dV_{eff}}{dr} ,
\end{equation}
where $V_{eff}$ is defined as
\begin{equation}\label{eq:veffec}
V_{eff} = -\frac{Q^2}{4 r^2}  - \frac{1}{2} \log r .
\end{equation}
This effective potential is sketched in Fig.~\ref{fig:rnbhtrajectories}.

The equation of motion Eq.~\ref{eq:Reom} is equivalent to a one dimensional particle moving in the potential Eq.~\ref{eq:veffec} subject to friction. There will be four qualitatively different types of trajectories, depending on the ``energy'' of the particle and the imposed boundary conditions: constant, bound, unbound, and monotonic.  (A similar effective potential analysis was performed in~\cite{ArkaniHamed:2007js} for a different class of black holes). Constant trajectories sit at critical points of the effective potential. Bound and unbound trajectories have turning points, with $r$ ranging between the turning point and $r = 0$ or $r \rightarrow \infty$ respectively. Monotonic trajectories encompass the entire range in $r$ from $0 < r < \infty$. 

The constant trajectories sit at the maximum of the effective potential over the entire range in $z$, and are equivalent to the solutions found above with $r' = r'' = 0$. Bound trajectories have their turning points to the left of the maximum of the effective potential, and unbound trajectories to the right. The bound and unbound trajectories obtained by evolving Eq.~\ref{eq:Reom} are uniquely determined by these turning points $r (0)$, where $r' (0) = 0$, and from Eq.~\ref{eq:Aadnrp}, $A (0)= 0$. Defining $a(r) = r'^2$, and changing coordinates from $z$ to $r$, the metric reduces to the static form of Eq.~\ref{eq:rnmetric}. The metric coefficient $a(r)$ goes to zero when $r'=0$, indicating that turning points in the motion correspond to the location of event horizons. Also, note that the mass parameter has not yet explicitly appeared in our analysis -- this will be determined by the turning points, since there is a one-to-one map between the horizon structure and the mass parameter for fixed charge. Taking $z \rightarrow i z$, regions where the spacelike and timelike coordinates switch are produced. All solutions to the equations of motion are oscillatory, since $r$ now evolves in the inverted potential, with the amplitude of oscillations specified by the turning point.

The full non-extremal black hole solutions can be produced by the procedure depicted in Fig.~\ref{fig:rnbhtrajectories}. First, choose a turning point to the right of the maximum, and evolve using the boundary conditions specified above to produce Region~I. Starting from the same turning point, analytically continue $z \rightarrow i z$, and evolve the Euclidean equations of motion to the second turning point (which will be to the left of the potential maximum) to produce Region~II. Analytically continuing back to a spacelike $z$ coordinate, evolve the equations of motion from the second turning point to $r = 0$, producing Region~III. The extremal solution, which has no regions where $z$ is timelike, corresponds to the trajectory that grazes the top of the effective potential, reaching the maximum only after an infinite span of $z$. This is the property of infinite proper distance to the horizon in the extremal geometry noted in Sec.~\ref{sec:geometry}. The monotonic trajectories have no turning points, and therefore no horizons, corresponding to the black hole geometries with $Q^2 > M^2$.

Thus, we see that the only new solution introduced by relaxing the requirement of asymptotic flatness in the uniqueness theorems is the compactification solution, the constant trajectory with $r'' = r' = 0$.

\section{The entropy of extremal black holes}
\label{sec:discussion}

The Einstein-Maxwell system gives rise to two static, spherically symmetric solutions described by the same set of conserved charges, but with different boundary conditions: the compactification solution and the extremal black hole. Either of these two solutions can be obtained from different parts of the non-extremal black hole (which is the unique solution specified by a mass and set of charges) using the limiting procedure described in Sec.~\ref{sec:extremal limit}. Thus, even at the level of classical geometry, there are subtleties in interpreting the limit of extremality. As we discussed in Sec.~\ref{sec:introduction}, semi-classical methods yield zero entropy for extremal black holes, while dual and string theory microstate counting predicts a non-zero entropy equal to $S = \pi \rho^2 / G$. The discontinuous extremal limit may shed some light on this apparent discrepancy, as we now discuss.

Before proceeding, it is instructive to review the argument given by~~\cite{Hawking:1994ii} for the vanishing entropy of extremal black holes. The semi-classical calculation seeks to evaluate the gravitational path integral in the saddle point approximation around Euclideanized black hole geometries~\cite{Gibbons:1976ue}. Euclideanizing a non-extremal black hole by sending $t_E = i t$ in Eq.~\ref{eq:rnmetric} where $r > r_+$ yields a manifold with topology $R^2 \times S^{2}$. The coordinates $\{r, t_E \}$ form a set of polar coordinates on the $R^2$ factor with the origin at $r_+$, and the periodicity $\beta$ of the angle $t_E$ set by imposing regularity (the absence of a conical singularity) at the origin. Regions of the Lorentzian manifold with $r < r_+$ are not part of the Euclidean solution. For an extremal black hole, since $r_+$ is infinitely far away from any point outside the horizon, this point is removed from the Euclidean manifold. The topology of the Euclidean extremal black hole is therefore $R \times S^1 \times S^2$.  Because the origin is removed, there will be no conical singularity for any choice for the periodicity of the Euclidean time.

The entropy is related to the Euclidean action by
\begin{equation}
S = \left( \beta \frac{d}{d\beta} - 1 \right) I_E .
\end{equation}
The Euclidean action for the non-extremal black hole receives ``boundary'' contributions from the vicinity of the origin (recall that there is the $S^2$ factor, which does not degenerate at the origin since $r = r_+$ here) that are independent of $\beta$, and contributions from the canonical action (the Euclideanized version of Eq.~\ref{eq:action}) that are proportional to $\beta$~\cite{Hawking:1994ii,Teitelboim:1994az}. The latter gives no contribution to the entropy and the former yields the Bekenstein-Hawking entropy $S = A / 4 G$. For the extremal black hole, because the origin is not part of the manifold (or, equivalently, because the periodicity of the the Euclidean time is not fixed), the contribution from the vicinity of the origin vanishes and the calculation indicates that the entropy of an extremal black hole is zero. 

We suggest that a possible resolution to the discrepancy between semiclassical methods and dual microstate counting is that the entropy of an extremal black hole does indeed vanish (agreeing with the semiclassical calculation), but the entropy of the corresponding compactification solution does not. That is, the dual microstate calculations describe the $AdS_2 \times S^2$ region, not the extremal black hole, whereas the semiclassical methods describe the extremal black hole, but not the compactification solution.

How plausible is this picture? One of the most robust methods of microstate counting, which is independent of the details of the underlying theory of quantum gravity, is the dual microstate counting of Strominger~\cite{Strominger:1997eq,Strominger:1998yg}. The original calculation was applied to 3 dimensional BTZ black holes~\cite{Strominger:1998yg}, but has subsequently been applied to Kerr-Newmann (charged, spinning) black holes in arbitrary dimensions~\cite{Guica:2008mu,Hartman:2008pb,Lu:2008jk}. The basic idea in each case is to exploit the fact that the isometries of the near horizon geometry in each of these extremal cases is $SL(2,R) \times U(1)$ to define a CFT on $AdS_2$. Key to these discussions were the asymptotic symmetries of global $AdS_2$, which all metric perturbations were required to respect, and the existence of a $U(1)$ (empty $AdS_2$ does not have the same properties). Cardy's formula~\cite{Cardy:1986ie} for the asymptotic growth of states in a CFT is then applied to obtain the entropy. 

This entropy is obtained from a calculation in global $AdS_2$, which has different boundary conditions than the original black hole, even though the extremal black hole only {\em locally} approximates $AdS_2 \times S^2$ at the horizon. From this perspective, it is clear that the state counting is done not for the original black hole, but from a dual holographic perspective in the near-horizon region. For this reason, we refer to these calculations as dual microstate counting. Because the entropy calculation applies to the global $AdS_2 \times S^2$ and not to the original black hole solution, it is not in obvious conflict with the semiclassical result that extremal black holes have vanishing entropy.

$AdS_2$ is special because it possesses two disconnected, timelike boundaries. An interesting proposed alternative explanation for the entropy of $AdS_2 \times S^2$ is that it arises as entanglement entropy (see \cite{Maldacena:2001kr,Ryu:2006ef,Ryu:2006bv} for further discussion of entanglement entropy in this context) between the degrees of freedom in region I and I' of the $AdS_2 \times S^{2}$~\cite{Azeyanagi:2007bj} in Fig.~\ref{fig:rnads}. For a nearly extremal black hole, the entanglement entropy arises from correlations in the very near vicinity of the event horizon~\cite{Frolov:1993ym,Das:2008sy}. It is precisely this region that becomes part of the compactification solution in the extremal limit, lending further support to the idea that the entropy in the extremal case is carried by the $AdS_2 \times S^2$ rather than by the extremal black hole. This also supports our suggested alternative non-holographic interpretation of the entropy arising from the bulk degrees of freedom of the $AdS_2$.

In summary, we propose that the entropy of extremal black holes might vanish whereas the entropy of $AdS_2 \times S^2$ does not. The entropy of the $AdS_2 \times S^2$ compactification solution does not vanish, as can be seen from the extremal microstate counting and entanglement entropy calculations. Further, the entropy is what one would obtain by a naive application of the Bekenstein--Hawking formula to an extremal black hole. This suggests that the semiclassical and dual microstate counting pictures could both be correct, as they are computing the entropy of two different spacetimes. It also suggests a non-holographic interpretation of the extremal entropy, which is carried by the compactification solution and associated with its bulk degrees of freedom.

This picture, while satisfying, leaves a few interesting puzzles. We have had little to say about string theory microstate counting for extremal black holes. What implications could the existence of two different solutions in the extremal limit have in this case? In addition, it would be interesting to construct a clear physical picture of the fate of the region between the inner and outer event horizons when classically attempting to assemble or destroy an extremal black hole. In both cases, the geometrical properties of the extremal limit may play an important role.

It can be argued that the extremal black hole geometry is not very physical. Quantum corrections to Einstein gravity will change the properties of the solutions, perhaps in a way that leads to a non-zero value for the entropy even from the standpoint of the semi-classical calculation (as suggested by Ref.~\cite{Horowitz:1996qd}). The implications of our proposal for this picture are unclear, but nevertheless our results can be viewed as a formal explanation of the discrepancy between various calculations for entropy of the idealized extremal black hole.

\begin{acknowledgments}
The authors wish to thank D. Anninos, T. Banks, A. Dabholkar, G. Horowitz, H. Ooguri, A. Strominger, and K. Vyas. Partial support for this research was provided by the U.S. Department of Energy and the Gordon and Betty Moore Foundation. L.R. is supported by NSF grant PHY-0556111.

\end{acknowledgments}

\bibliography{blackholeentropy}

\end{document}